\documentclass[traditabstract]{aa}

\usepackage{graphicx}
\usepackage{rotating}
\usepackage{txfonts}

\begin{document}

\title{Herschel PACS and SPIRE Observations of TWA brown dwarf discs\thanks{{\it Herschel} is an ESA space observatory with science instruments provided by European-led Principal Investigator consortia and with important participation from NASA.} }

\titlerunning{{\it Herschel} Observations of TWA brown dwarf discs}

\author{B. Riaz
              \inst{1}
               \& J. E. Gizis
              \inst{2}}

\institute{Centre for Astrophysics Research, Science \& Technology Research Institute, University of Hertfordshire, Hatfield, AL10 9AB, UK \\
\and Department of Physics and Astronomy, University of Delaware, Newark, DE 19716, USA }

\date{Recieved --- ; Accepted ---}

\abstract{ We present {\it Herschel} SPIRE observations for the TW Hydrae association (TWA) brown dwarf discs SSSPM J1102-3431 (SS1102) and 2MASSW J1207334-393254 (2M1207). Both discs are undetected in the SPIRE 200-500$\mu$m bands. We have also analyzed the archival PACS data and find no detection for either source in the 160$\mu$m band. Based on radiative transfer modeling, we estimate an upper limit to the disc mass for both sources of 0.1 $M_{Jup}$. The lack of detection in the SPIRE bands could be due to a paucity of millimeter sized dust grains in the 2M1207 and SS1102 discs. We also report a non-detection for the brown dwarf 2MASS J1139511-315921 (2M1139) in the PACS 70 and 160$\mu$m bands. We have argued for the presence of a warm debris disc around 2M1139, based on an excess emission observed at 24$\mu$m. The mid-infrared colors for 2M1139 are similar to the transition discs in the Taurus and Ophuichus regions. A comparison of the brown dwarf disc masses over a $\sim$1-10 Myr age interval suggests a decline in the disc mass with the age of the system.  }


\maketitle

\section{Introduction} 

The TW Hydrae association (TWA) has been under observation since its identification by Kastner et al. (1997). This young (10$^{+5}_{-2}$ Myr; Barrado y Navascu\'{e}s 2006) and nearby ($\sim$53pc, Gizis et al. 2007) association now has more than 30 confirmed members, in both stellar and sub-stellar regimes, with a few members displaying signatures of being in the T Tauri phase. Given its proximity and a critical intermediate age when most discs begin to show a transition to the debris phase, TWA is well-suited for the study of disc evolution and planet formation. Using ground-based mid-infrared observations, Jayawardhana et al. (1999) and Weinberger et al. (2004) were able to detect excess thermal emission at 10, 12 and 18 $\mu$m for four TWA members (TW Hya, Hen 3-600, HD 98800A and HR 4796A). However, most of the other members do not show any evidence for circumstellar dust at these wavelengths, indicating cleansing of dust in the inner disc regions on a time scale of $\leq$10 Myr. A {\it Spitzer} survey by Low et al. (2005) using the sensitive MIPS instrument  revealed the presence of cold debris discs around two TWA members, TWA 7 and 13, at 70 $\mu$m, that remained undetected by {\it IRAS} and other previous surveys. Among the brown dwarf members of TWA, excess emission has been detected at mid- and far-infrared wavelengths for 2MASSW J1207334-393254 (2M1207) and SSSPM J1102-3431 (SS1102), while a small 24 $\mu$m excess has been reported for 2MASSW J1139511-315921 (2M1139), suggesting the presence of a transition disc around this source (Riaz et al. 2006; Riaz \& Gizis 2008). By transition discs, we imply a disc with an inner opacity hole. The inner disc in such cases may have optically thin dust, or could be completely devoid of circumstellar material. Due to reduced opacity in the inner disc regions, transition discs exhibit photospheric emission shortward of a certain mid-infrared wavelength, depending on the size of the inner disc hole. Based on the overall disc fractions and the strength in the mid-infrared excess emission, Riaz \& Gizis (2008) suggested longer disc dissipation timescales for the brown dwarfs in TWA. This association has proven to be an important field to study the formation and evolutionary differences, if any, in stellar and sub-stellar objects, and their disc evolution time scales.


We present ESA Herschel Space Observatory (Pilbratt et al.\ 2010) SPIRE (Griffin et al.\ 2010) observations at 200-500$\mu$m for SS1102 and 2M1207, and PACS (Poglitsch et al. 2010) 70 and 160$\mu$m observations for 2M1139. Observations at far-infrared wavelengths for SS1102 and 2M1207 can help constrain the model fits and obtain better estimates for the disc mass and outer disc radius, while PACS observations of 2M1139 are important to probe the presence of cold dust material for this transition disc. Our Herschel observations are presented in Section \S\ref{observations}. Results from the radiative transfer modeling are presented in Section \S\ref{models}. Section \S\ref{discussion} provides a comparison of the TWA brown dwarf disc masses with the higher mass stars in this association and with the younger brown dwarf discs in Taurus, along with a discussion on the nature of the transition disc around 2M1139.

\section{Observations and Data Analysis}
\label{observations}

We obtained SPIRE photometric observations for SS1102 and 2M1207 in the 250, 350 and 500$\mu$m bands, using the small map mode (ObsID: 1342234825/1342223257). The small map mode covers a 5$^{\prime}$ diameter circle, with a fixed scan angle of 42.4$\degr$ and a fixed scan speed of 30$^{\prime\prime}$s$^{-1}$. We requested a setup of 25 repetitions, resulting in a total on-source integration time of 925s. For 2M1139, PACS photometric observations were obtained in the scan map mode in the blue (70$\mu$m) and the red (160$\mu$m) filters. Two different scan map angles of 45$\degr$ and 63$\degr$ were used (ObsID: 1342224188/89). Both scans were obtained at a medium scan speed of 20$^{\prime\prime}$/s, with a cross scan step of 5$^{\prime\prime}$ and scan leg length of 3$^{\prime}$. The number of scan legs was 8 for the 63$\degr$ scan, and 4 for the 45$\degr$ scan. 

Data analysis for all sources was performed using the rectangular sky photometry task provided in the Herschel Interactive Processing Environment (HIPE version 8.0.1). We have worked on the final pipeline processed data (the `Level 2' products). The rectangular photometry task applies aperture photometry for a circular target aperture and a rectangular sky aperture. For 2M1139, photometry was performed on the mosaic of the two scan maps. The sky intensity was estimated using the median sky estimation algorithm. Since the sky intensity could vary with location, we selected 4 different background regions around the target (each of 30x30 pixel area) and measured the sky-subtracted source flux for each sky region. The final source flux and the flux error is the mean and the standard deviation of these 4 flux measurements. The flux errors are consistent with the values obtained at the source location from the `error' image provided in the Level 2 products. 

\subsection{SS1102}

\begin{figure*}
\centering
  \includegraphics[width=8cm, angle=0]{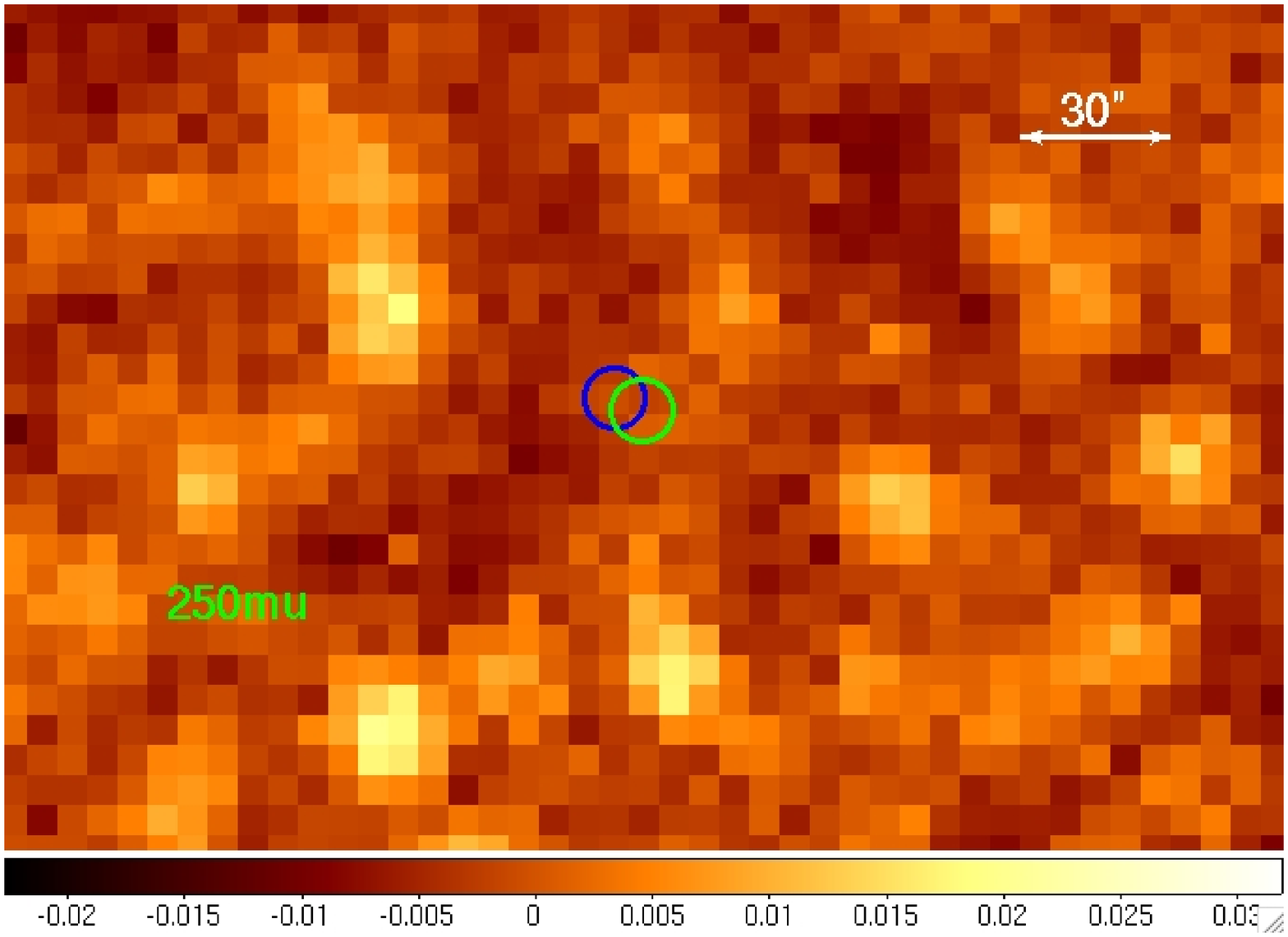} \hspace{0.1in}
  \includegraphics[width=8cm, angle=0]{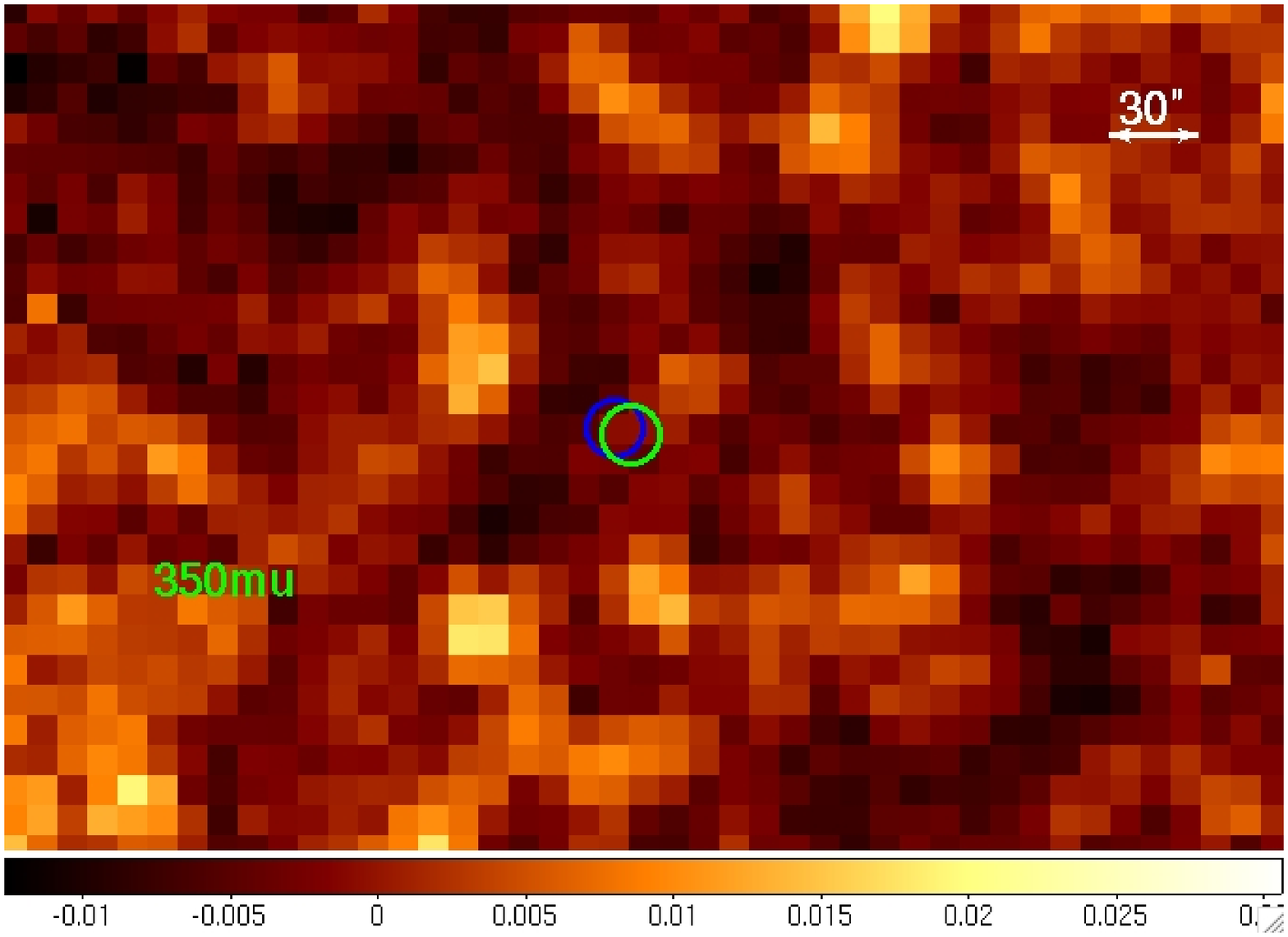}
  \includegraphics[width=8cm, angle=0]{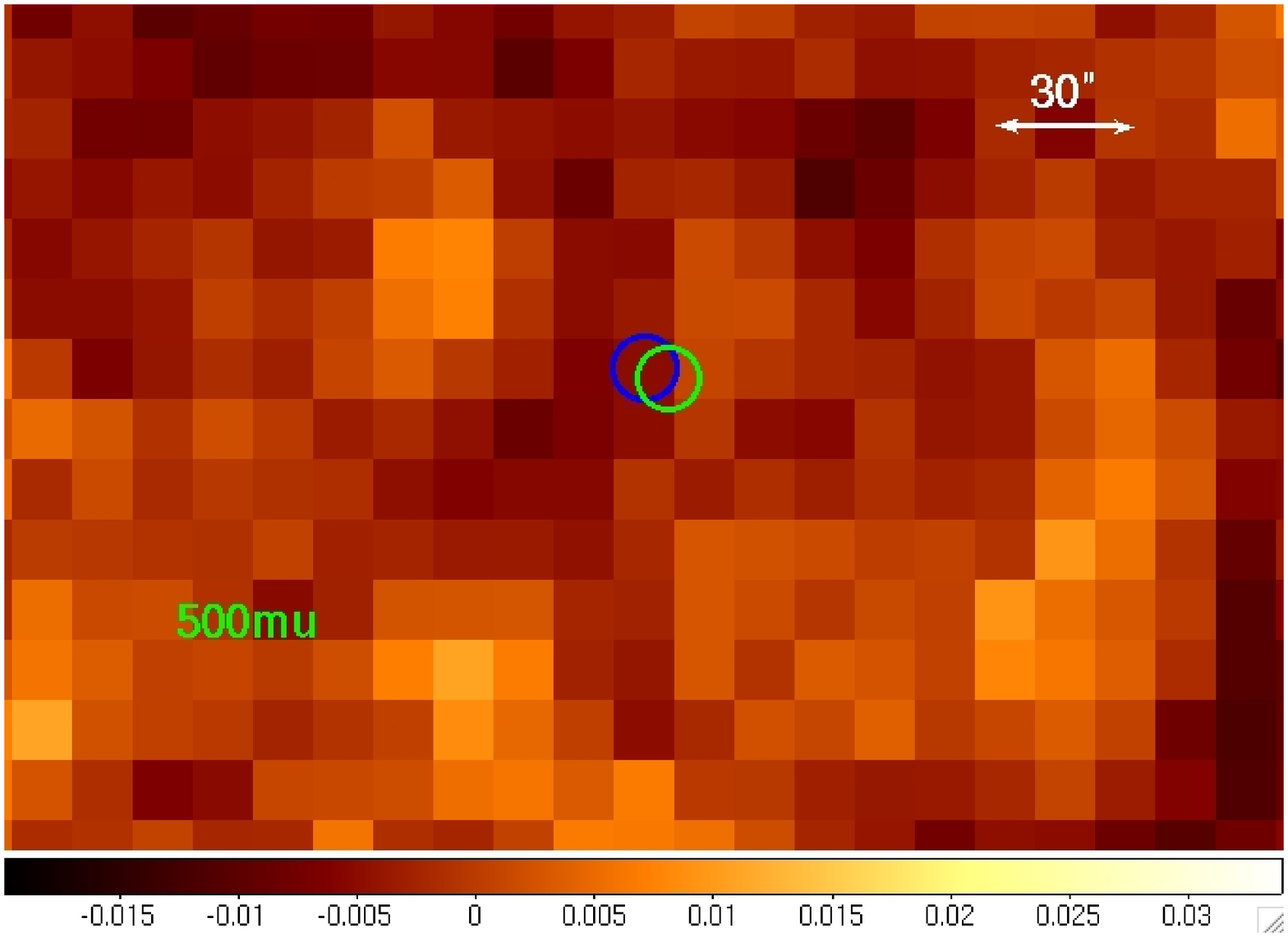}    
 \caption{The SPIRE images for SS1102. Blue circle marks the actual target position, green circle marks the nominal position. The image scale for 30$\arcsec$ is shown in the top right. North is up, East is to the left.   }
  \label{image-1102}
\end{figure*}

SS1102 is undetected in all SPIRE bands (Fig.~\ref{image-1102}). The positional accuracy as determined from the actual ra, dec and the raNominal, decNominal in the metadata is $\sim$1 pixel (or $\sim$6$^{\prime\prime}$ in the 250$\mu$m band). We have calculated the upper limits using a pixel radius of 2 pixels centered at the nominal position of SS1102. Harvey et al. (2012) have reported a detection for SS1102 in the PACS 160$\mu$m band, at a $\sim$7 mJy level. We have analyzed their PACS data, and find no source detection at the nominal or the actual position of the target (Fig.~\ref{image-160mu}). The flux level at the target location is $\sim$0.1-0.2 mJy, and is the same as found in some of the confusion noise dominated regions seen in the image. The offset between the actual and the nominal positions in the metadata is $\sim$1.4$^{\prime\prime}$, or 3.5 pixels. However, there is no source detection at a 2-$\sigma$ level even if the offset is taken into account. The 160$\mu$m flux measurement for SS1102 reported by Harvey et al. (2012) is thus incorrect. Our analysis of SS1102 in the PACS 70$\mu$m band, however, is consistent with the measurement reported by Harvey et al. (2012). Table~\ref{fluxes} lists the updated fluxes for SS1102.

\subsection{2M1207}

We had reported in Riaz et al. (2012a) a detection for 2M1207 in the SPIRE bands of 250 and 350$\mu$m, but a comparison now with the PACS images indicates this to be a misidentification. The revised SPIRE fluxes (upper limits) for 2M1207 have been presented in Riaz et al. (2012b). As noted in that work, there is a possibility of having an unresolved source in the SPIRE bands, given the large positional offset of $\sim$14$^{\prime\prime}$. However, the contamination from the nearby bright object is likely to dominate the composite photometry in such a scenario. The SPIRE upper limits have been determined by measuring the emission at the nominal position for 2M1207, without considering the offset. 

\begin{figure*}
\centering
  \includegraphics[width=8cm, angle=0]{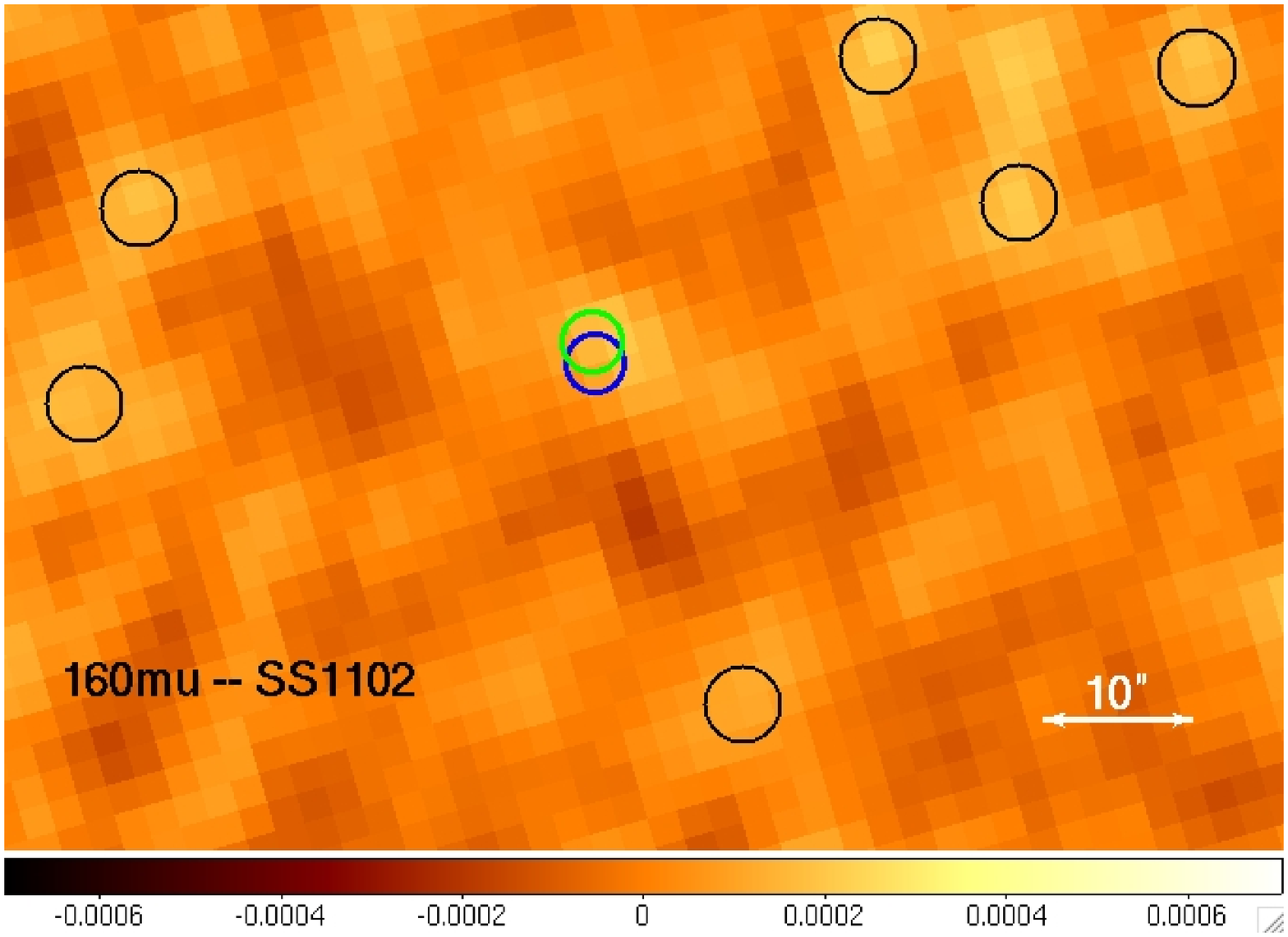} \hspace{0.1in}
  \includegraphics[width=8cm, angle=0]{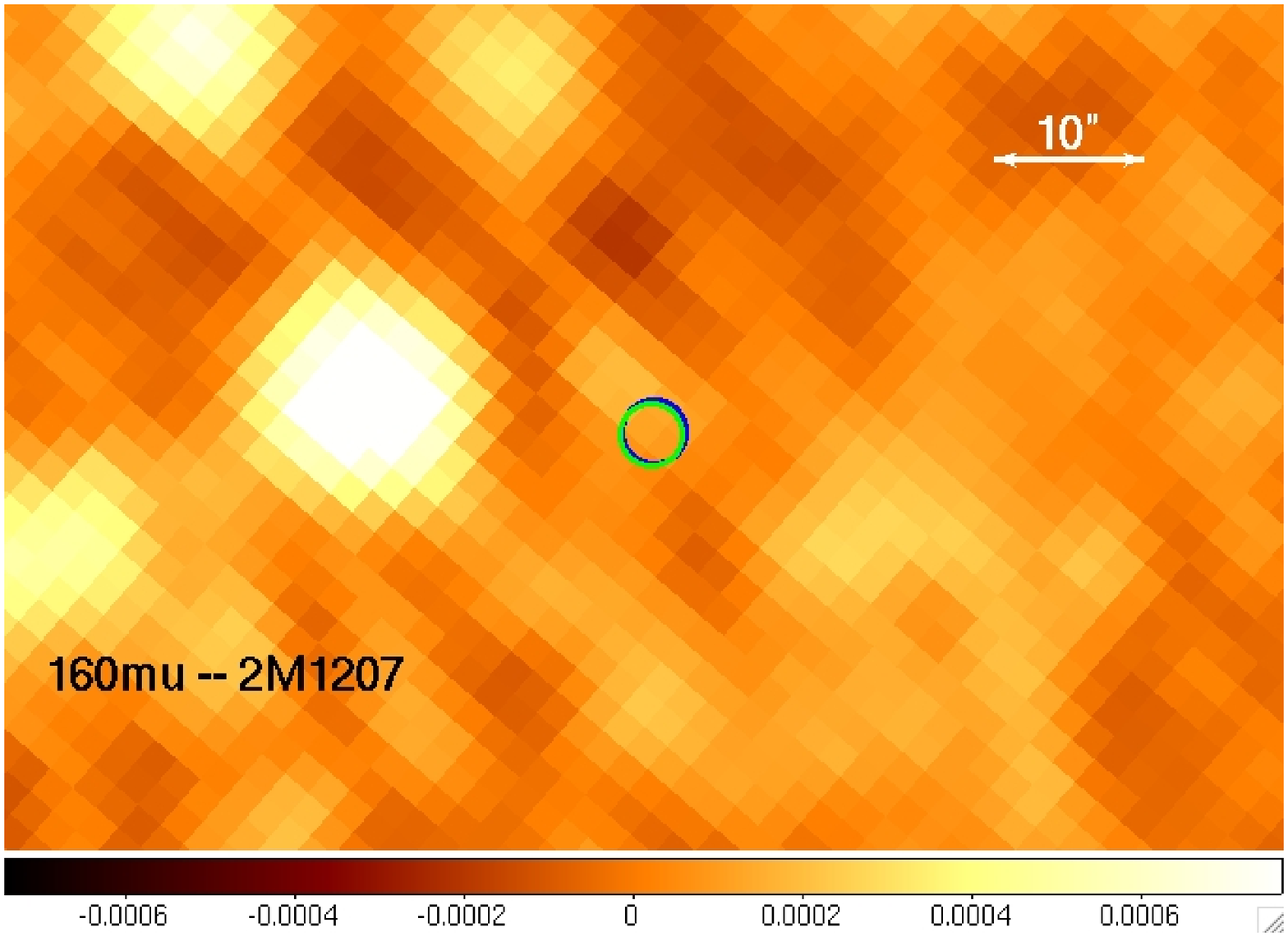} 
 \caption{The PACS 160$\mu$m images for SS1102 (left) and 2M1207 (right). Blue circle marks the actual target position, green circle marks the nominal position. For the SS1102 image, black circles indicate some of the confusion noise dominated regions with similar flux levels as observed at the nominal target position. The image scale for 10$\arcsec$ is shown in the top right. North is up, East is to the left. }
  \label{image-160mu}
\end{figure*}

We have also analyzed the PACS 70 and 160$\mu$m observations for 2M1207. Our 70$\mu$m measurement is consistent with the flux reported by Harvey et al. (2012). In the 160$\mu$m image, there is some diffuse emission seen at a distance of 2 pixels ($\sim$0.8$^{\prime\prime}$) from the nominal position of 2M1207 (Fig.~\ref{image-160mu}). This diffuse emission is at a flux level of $\sim$0.2mJy, similar to that observed at the location of SS1102 in the 160$\mu$m data. We can consider this flux level as the expected confusion noise in the 160$\mu$m band, based on which the detection of both 2M1207 and SS1102 at 160$\mu$m can be rejected.

\subsection{2M1139}

\begin{figure*}
\centering
  \includegraphics[width=8cm, angle=0]{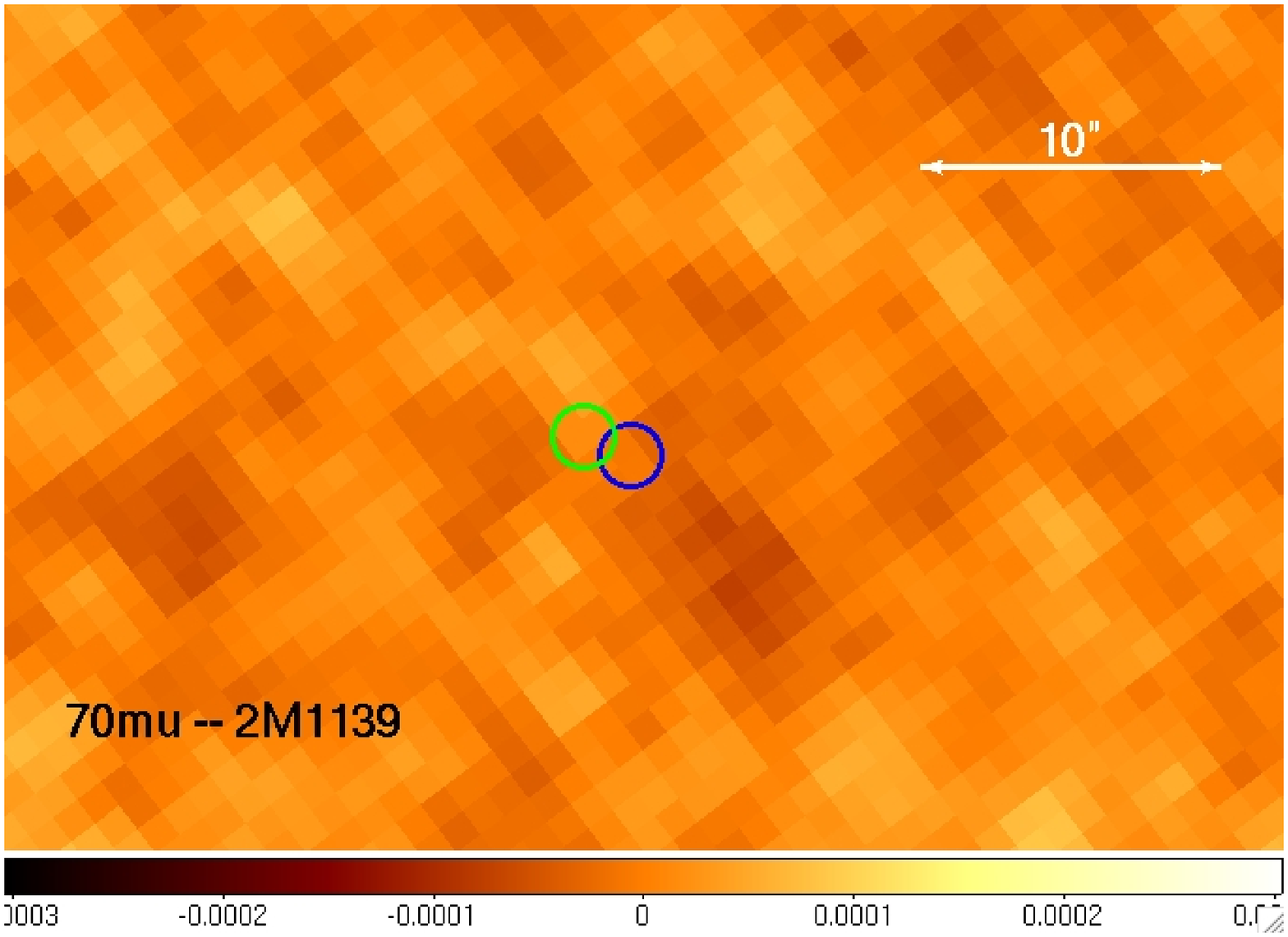} \hspace{0.1in}
  \includegraphics[width=8cm, angle=0]{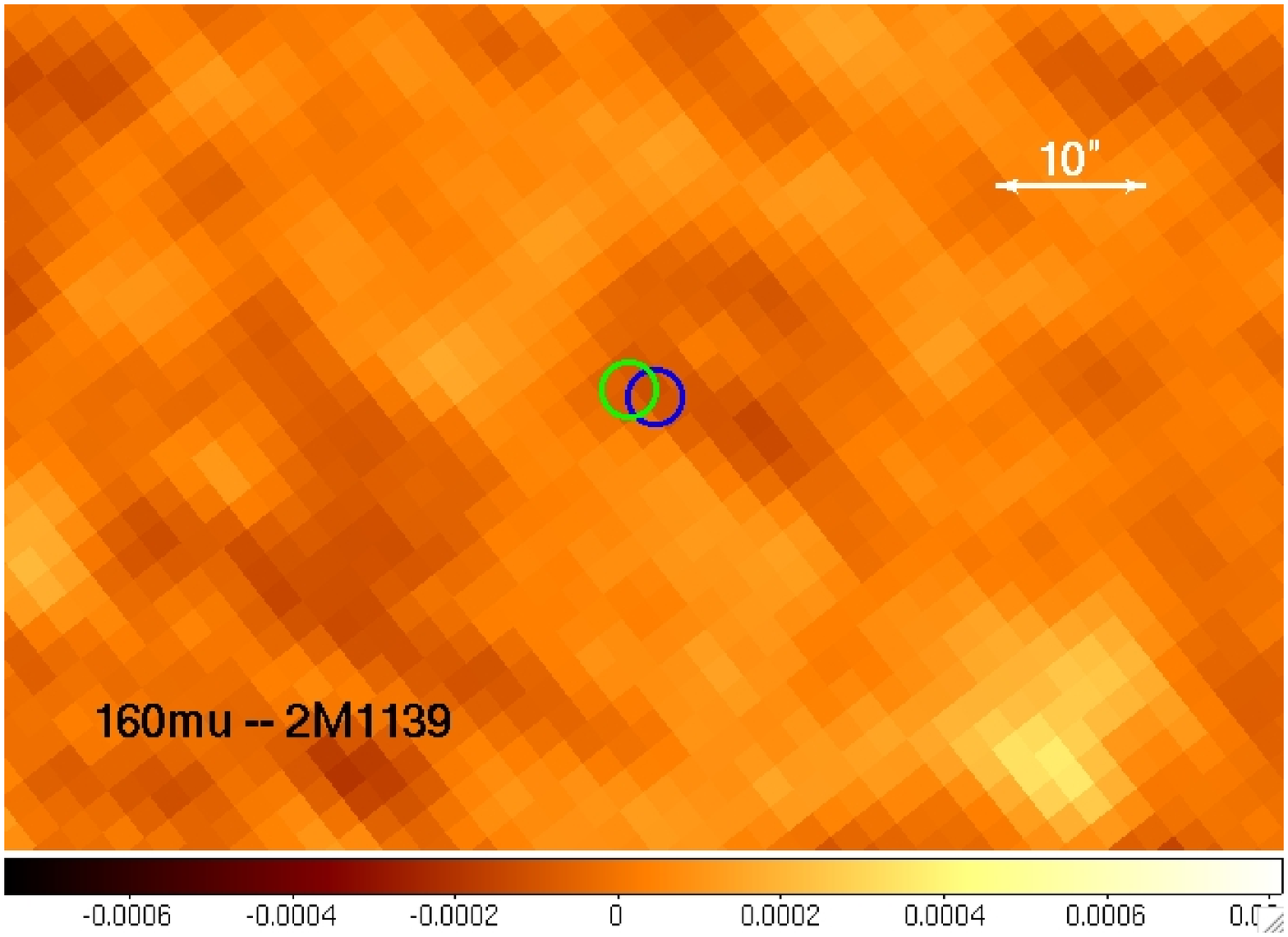}
 \caption{PACS images for 2M1139. Blue circle marks the actual target position, green circle marks the nominal position. The image scale for 10$\arcsec$ is shown in the top right. North is up, East is to the left.  }
  \label{image-1139}
\end{figure*}

2M1139 is undetected in the PACS bands of 70 and 160$\mu$m (Fig.~\ref{image-1139}). The position uncertainty is $\sim$0.5 pixel ($\sim$0.2$^{\prime\prime}$). We have determined the upper limits by measuring the flux at the nominal position of the target. We note that the {\it Spitzer} 70$\mu$m upper limit reported in Riaz et al. (2009) is much higher than the PACS 70$\mu$m measurement, which can be expected given the higher confusion noise in the {\it Spitzer} MIPS bands compared to PACS.

\section{Disk Modeling}
\label{models}

\begin{figure*}
\centering
 \includegraphics[width=80mm]{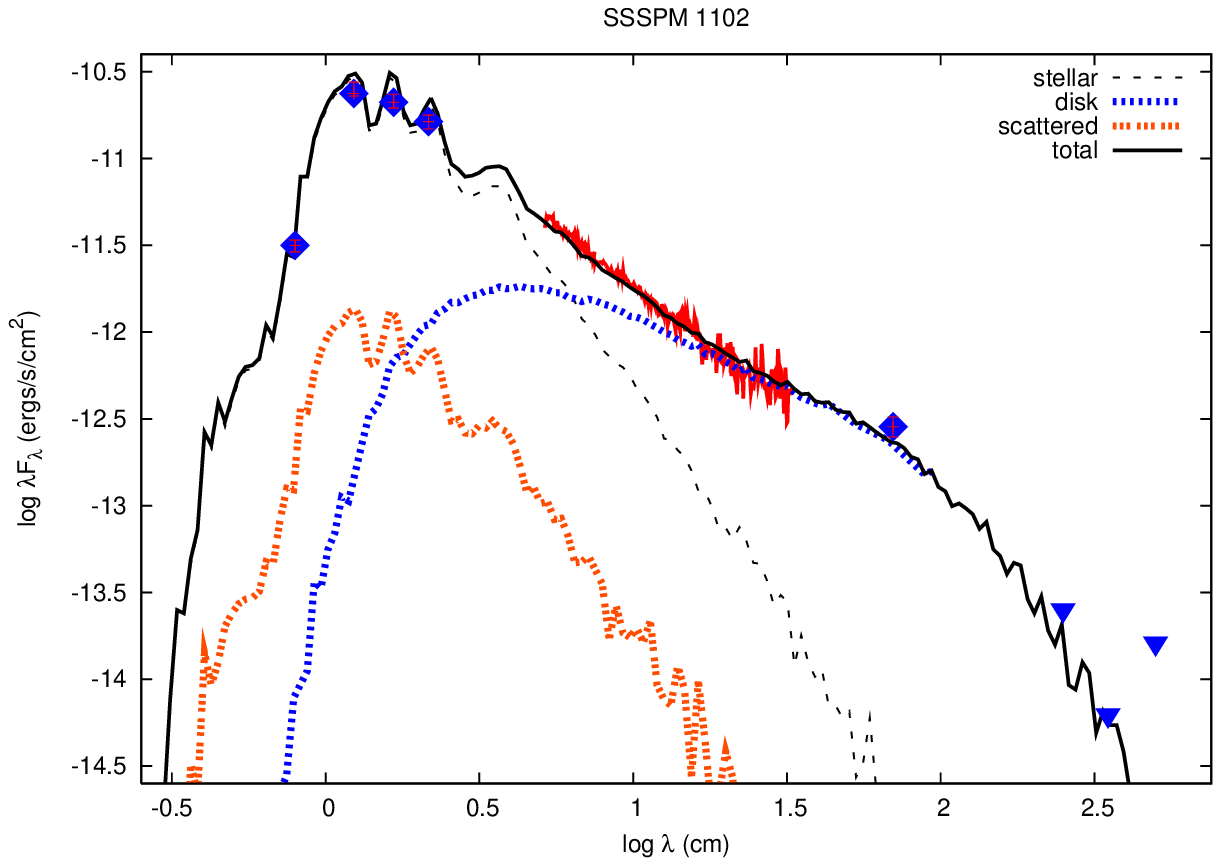} 
\includegraphics[width=80mm]{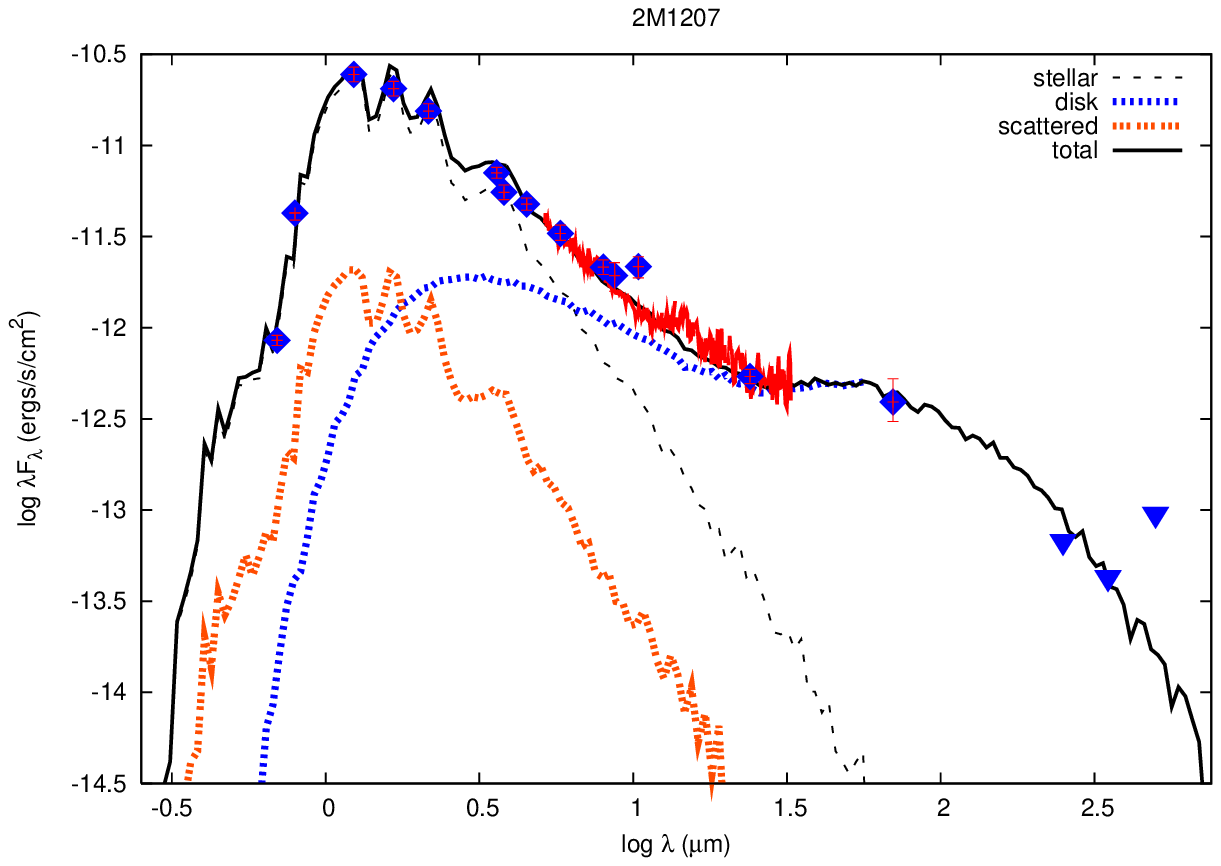} 
\includegraphics[width=80mm]{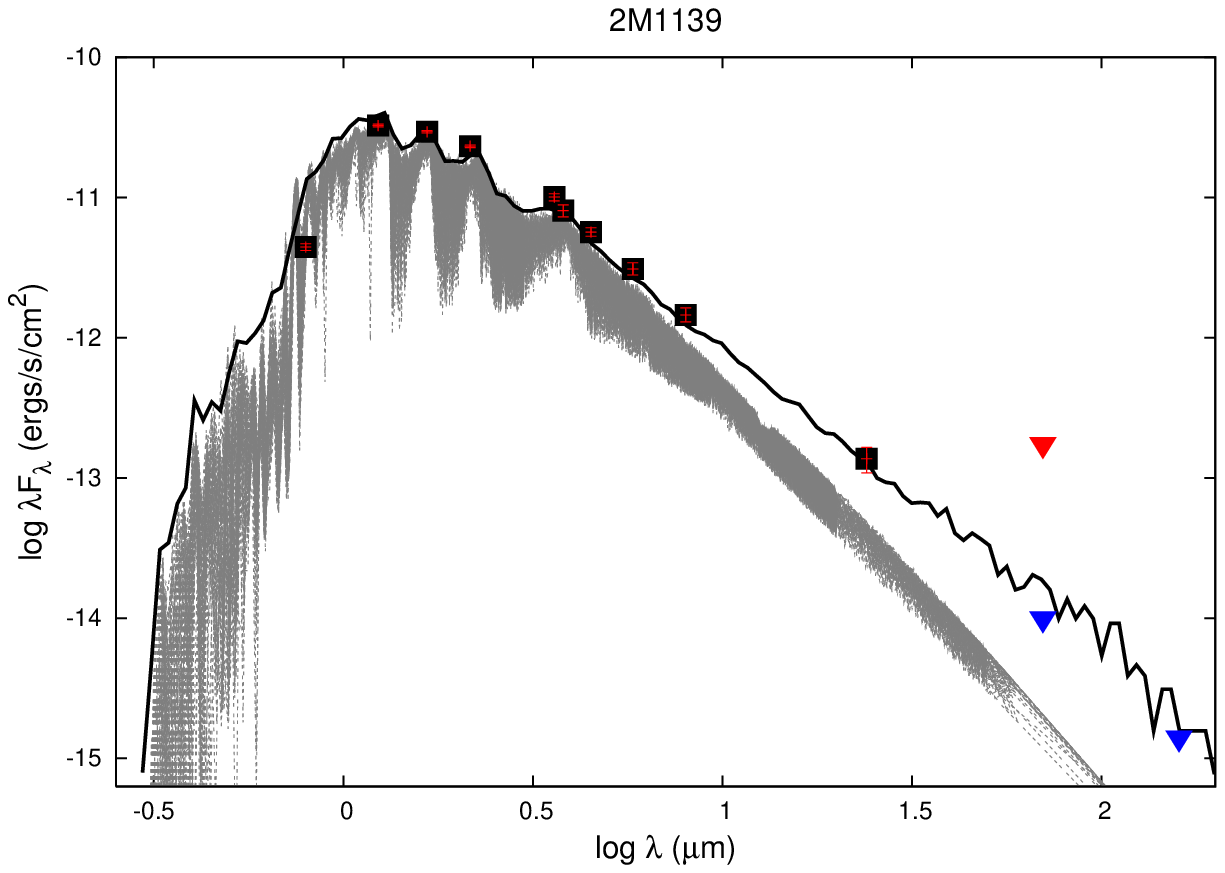} 
    \caption{The best model fits for SS1102, 2M1207 and 2M1139 ({\it bottom}). Also shown is the contribution from the disc (blue) and the stellar photosphere (grey). The Spitzer/IRS spectrum is shown in red. Blue points are the {\it Herschel} upper limits. For 2M1139, the red point marks the {\it Spitzer} 70$\mu$m upper limit. }
   \label{model}
 \end{figure*}

Figure~\ref{model} shows the model fits for SS1102 and 2M1207. The fits are based on 2-D radiative transfer modeling of near-infrared to sub-millimeter data, using the code by Whitney et al. (2003). Detailed discussion on the fitting procedure and the variations in the model SEDs is provided in Riaz \& Gizis (2007). Here we have provided a brief description of the best model-fits obtained, based on the lowest $\chi^{2}$ value. There are mainly six parameters related to the disc emission which can be varied to obtain a good fit (Table~\ref{disc}). The disc scale height varies as $h=h_{0}(\varpi / R_{*})^{\beta}$, where $h_{0}$ is the scale height at $R_{*}$ and $\beta$ is the flaring power. Both 2M1207 and SS1102 show flat disc structures, and values for $\beta$ of $\sim$1.1-1.2, and $h_{0}$ of $\sim$0.3 provide a good fit. We have used the large grain model ($a_{max}$=1mm) in the disc midplane and the medium sized grains ($a_{max}\sim$ 1 $\mu$m) in the disc atmosphere. Since both discs show no silicate emission at 10$\mu$m (Riaz \& Gizis 2008), grains larger than the ISM are required to fit the flat silicate feature. Further details on the variations in the model SEDs for the different grain models are discussed in Riaz \& Gizis (2007). Due to binning of photons in the models, there are a total of 10 viewing angles available. For 2M1207, a good fit was obtained using close to edge-on inclination angles between 57$\degr$ and 69$\degr$, whereas SS1102 is modeled well using intermediate angles of 53$\degr$-63$\degr$. We have varied the inner disc radius $R_{min}$ in multiples of $R_{sub}$, which is the dust sublimation radius. We have obtained good fits for $R_{min}$ between 1 and 3$R_{sub}$, with a peak at 1$R_{sub}$ (based on a $\chi^{2}$ value comparison). This implies an absence of an inner hole in the disc since it would have to be larger than the dust sublimation radius. Using the sub-mm upper limits, the best model fit for both 2M1207 and SS1102 is for an outer disc radius, $R_{max}$, of 50 AU and an upper limit to the disc mass of 0.1 $M_{Jup}$. A 100 AU model for a mass of 0.5 $M_{Jup}$ is also a good fit, though at a slightly higher $\chi^{2}$ value than the best fit. The 10 and 30 AU models are not a good fit to the longer part of the IRS spectrum for wavelengths $>$10$\mu$m. The optical depth is too large for these models even if the disc mass is reduced to 0.01 $M_{Jup}$, and misses the 70-350$\mu$m points. Increasing the disc mass for the same radius increases the optical depth beyond $\sim$30$\mu$m, while increasing the radius for the same disc mass reduces the optical depth longward of $\sim$30$\mu$m. The best fits for 2M1207 and SS1102 are the same as provided in Riaz \& Gizis (2008). We refer the readers to that paper for details on the rest of the fitting parameters. 

We have also attempted to model the 24$\mu$m excess along with the 70 and 160$\mu$m upper limits for 2M1139 (Fig.~\ref{model}). We have used the large grain model ($a_{max}$=1mm) in both the disc midplane and the atmosphere to fit this source. The best fit is obtained for a disc mass of 1E-7 $M_{\sun}$ and an outer disc radius of 100 AU. Since this is a transition disc with an inner opacity hole, a larger value of $R_{min}$$\sim$7 $R_{sub}$ provides a good fit. The parameter values listed in Table~\ref{disc} for 2M1139 model fit should be considered with caution. These or any of the other disc parameters cannot be properly constrained, considering that we have only one excess point and two upper limits. This is particularly for the case of the disc inclination angle and the outer disc radius. The model fit for 2M1139 mainly indicates that there may be a small mass of warm dust located at $\sim$1 AU distance from the central brown dwarf.

\section{Discussion}
\label{discussion}

\begin{figure*}
\centering
  \includegraphics[width=10cm, angle=270]{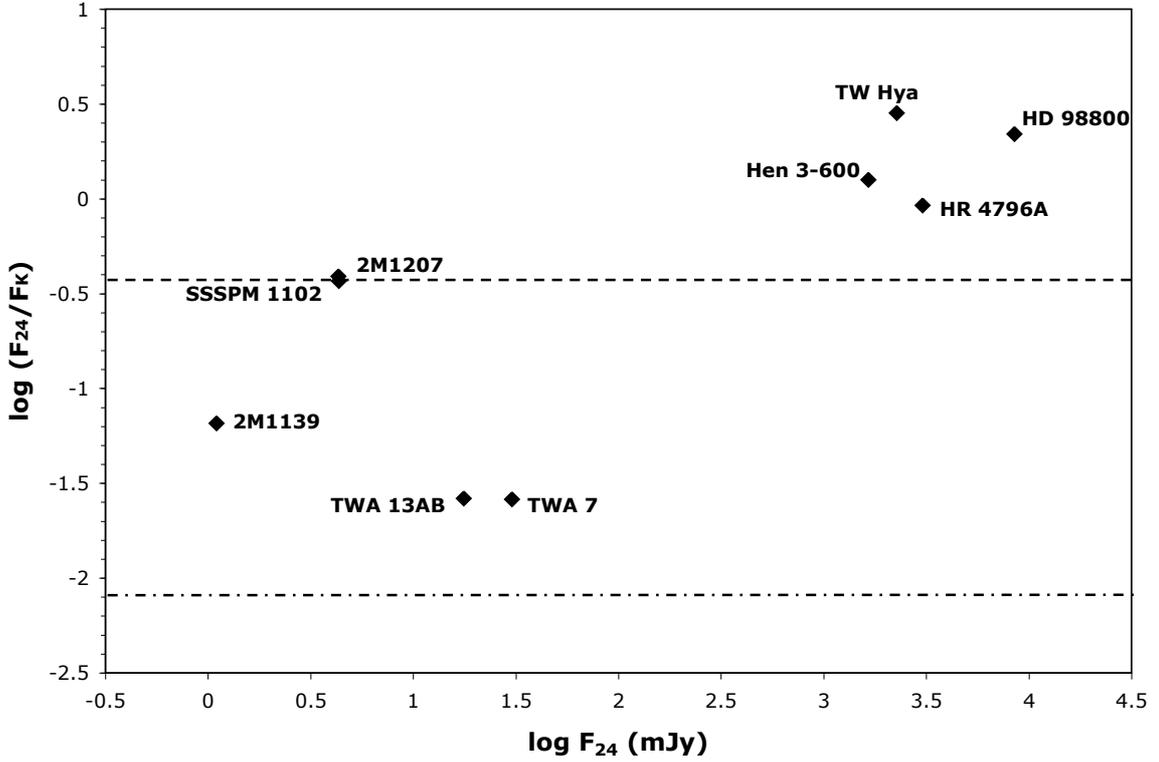}
 \caption{The extent of excess emission at 24 $\mu$m for the TWA stars and brown dwarfs. The dot-dashed line represents the limit of $F_{24}/F_{K}$ under the assumption that both bands lie on the Rayleigh-Jeans tail of the stellar spectrum. The dashed line represents a geometrically thin, optically thick flat disc with a spectral slope $\lambda F_{\lambda} \propto \lambda^{-4/3}$. }
  \label{excess}
\end{figure*}

\begin{figure*}
\centering
  \includegraphics[width=14cm, angle=0]{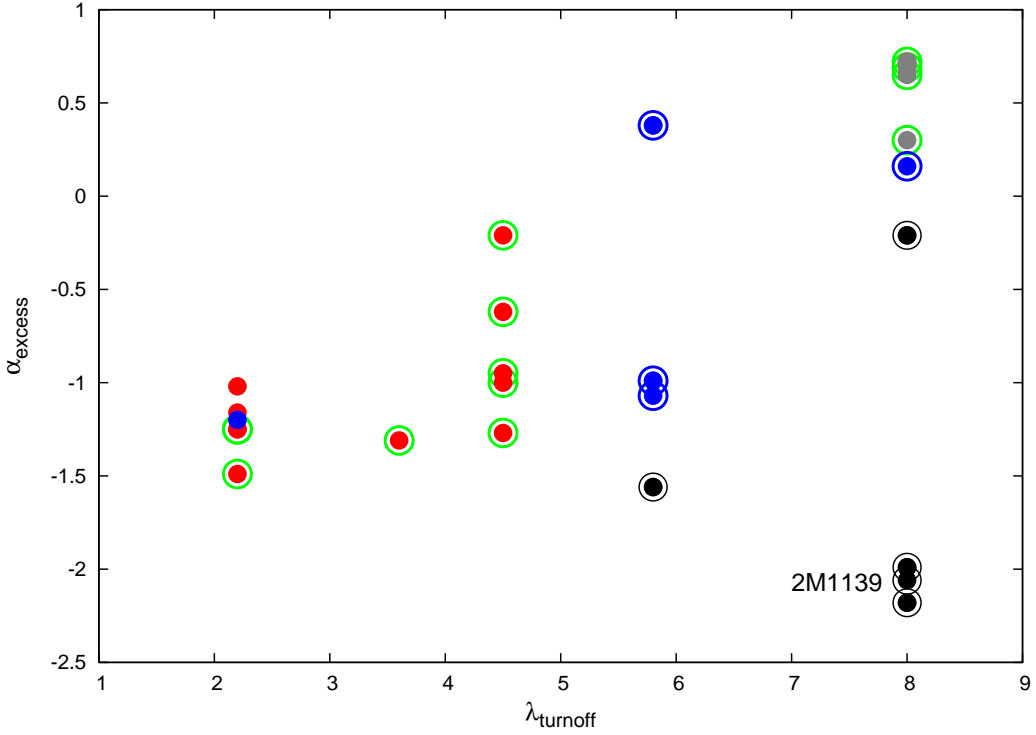}
 \caption{A $\alpha_{excess}$ vs. $\lambda_{turnoff}$ diagram showing 2M1139 and the transition disc samples in Ophuichus and Taurus. Transition discs with inner holes due to grain growth/dust settling are indicated by red symbol, due to photoevaporation by blue, and due to giant planet formation by grey. Debris discs are indicated by black filled symbol. Open green circle indicates objects with accretion rates of 1E-11 to 1E-9 $M_{\sun}$ yr$^{-1}$ and $L_{D}/L_{*}$ between 1E-3 to 1E-2. Blue open circles indicate non-accretiors ($<$1E-11 $M_{\sun}$ yr$^{-1}$) with $L_{D}/L_{*}$ between 1E-3 to 1E-2.. Black open circles indicate non-accretiors ($<$1E-11 $M_{\sun}$ yr$^{-1}$) with weak discs ($L_{D}/L_{*}$ $<$ 1E-3).  }
  \label{turnoff}
\end{figure*}

Figure~\ref{excess} shows the extent of excess emission at 24 $\mu$m for the TWA stars and brown dwarfs. The dot-dashed line represents the limit of $F_{24}/F_{K}$ under the assumption that both bands lie on the Rayleigh-Jeans tail of the stellar spectrum. The $F_{24}$-to-$F_{K}$ ratio for 2M1139 is higher than TWA 7 or TWA 13, indicating a warmer debris disc around this brown dwarf compared to the cold 80 K disc around TWA 7 or the 65 K disc around TWA 13AB. The dashed line represents a geometrically thin, optically thick flat disc with a spectral slope $\lambda F_{\lambda} \propto \lambda^{-4/3}$. 2M1207 and SS1102 lie just at the dashed line, indicating optically thick flat discs around these two brown dwarfs. The primordial nature of 2M1207 and SS1102 discs has also been argued in Riaz et al. (2012c), based on the strength in the excess emission observed in the {\it Spitzer} IRAC bands and the IRS spectrum. That these discs are optically thick can also be inferred from the $K_{s}$ to 24$\mu$m slope, $\alpha$, of the SED. We can consider a boundary of $\alpha$ = -2.2 to separate the Class II and Class III sources (e.g., Luhman et al. 2010). 2M1207 and SS1102 have $\alpha$ of -1.38 and -1.42, respectively, consistent with being Class II systems, while 2M1139 has $\alpha$ of -2.17 and is just above the Class III boundary. Large negative spectral slope values can be expected for transition discs with photospheric emission shortward of 24$\mu$m due to inner opacity holes (e.g., Cieza et al. 2010). Transition discs can be better characterized in a $\alpha_{excess}$ vs. $\lambda_{turnoff}$ diagram (Fig.~\ref{turnoff}), where $\lambda_{turnoff}$ is the longest wavelength at which the observed emission is photospheric, and $\alpha_{excess}$ is the spectral slope between $\lambda_{turnoff}$ and 24$\mu$m (e.g., Cieza et al. 2010). 2M1139 has $\lambda_{turnoff}$ of 8$\mu$m and $\alpha_{excess}$ of -2.1. The parameter $\lambda_{turnoff}$ is a good indicator of the size of the inner opacity hole in a transition disc. For a sub-stellar source, $\lambda_{turnoff}$ of 8$\mu$m implies an inner hole of $\sim$1 AU (e.g., Muzerolle et al. 2006; Cieza et al. 2010). We have included in Fig.~\ref{turnoff} the sample of young transition discs in the Ophichus and Taurus regions presented in Cieza et al. (2010; 2012). A large negative value for $\alpha_{excess}$ indicates a debris disc or a continuous disc that has undergone significant grain growth and dust settling, compared to large positive values that indicate an outer optically thick disc with a sharp inner cut due to e.g., the presence of a companion in the inner opacity hole. The location of 2M1139 in Fig.~\ref{turnoff} is consistent with being a warm debris disc. It has a low fractional disc luminosity ($L_{D}/L_{*}$$<$10$^{-4}$) and is a non-accretor ($<$ 10$^{-11}$$M_{\sun}$yr$^{-1}$), similar to the debris disc candidates in the Ophichus and Taurus regions. 

\begin{figure*}
\centering
  \includegraphics[width=14cm, angle=0]{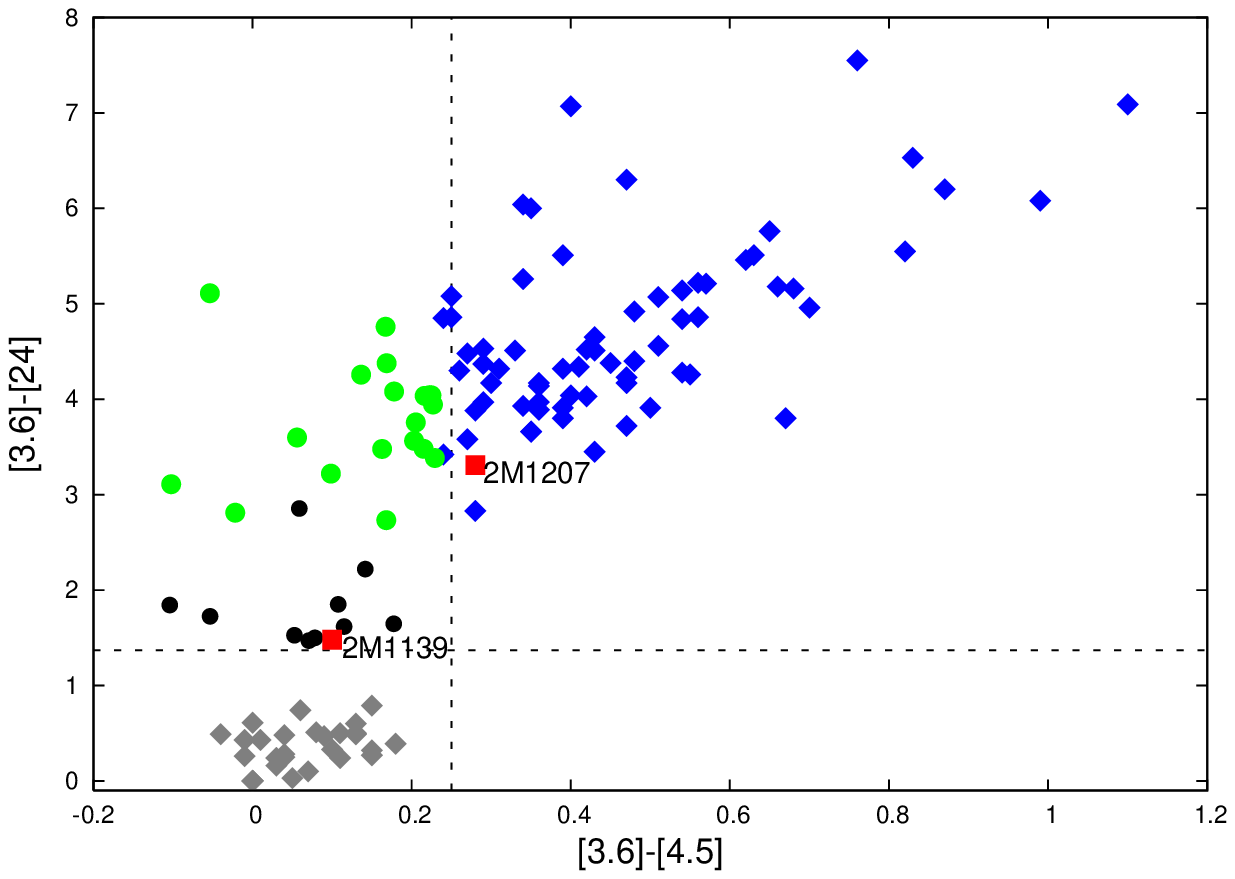}
  \includegraphics[width=14cm, angle=0]{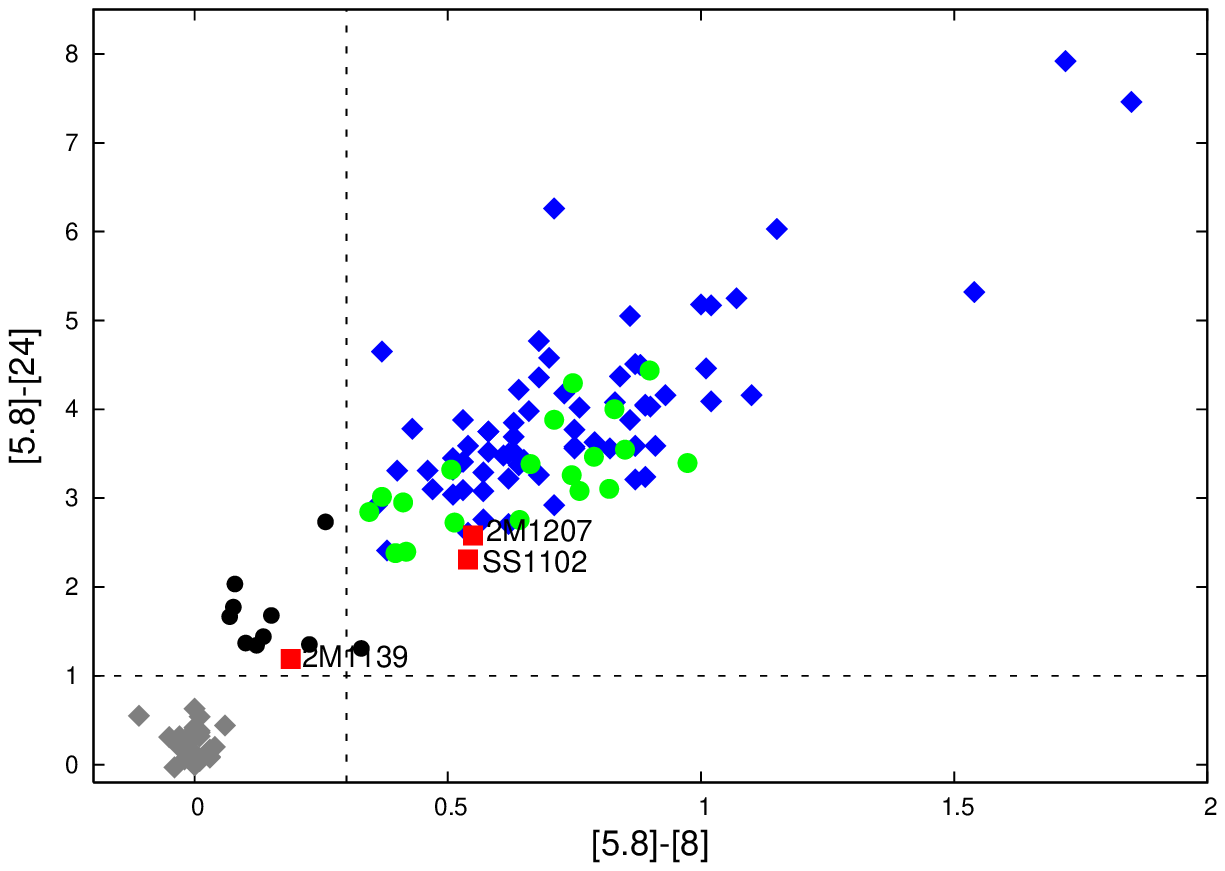}  
 \caption{Mid-IR color-color diagram for TWA brown dwarf discs. Also included for comparison are the discless (grey) and Class II (blue) sources in Taurus, and the transition disc sample in Ophuichus and Taurus. Green symbol indicates the transition discs with inner disc evacuation due to grain growth/dust settling. Black symbols indicate the debris discs.  }
  \label{colors}
\end{figure*}

Fig.~\ref{colors} compares the mid-IR colors for the TWA brown dwarf discs with the Class II and the discless sources in Taurus (from Luhman et al. 2010) and the transition disc samples in Ophichus and Taurus from Cieza et al. (2010; 2012). The fluxes at 3.6 and 4.5$\mu$m are not known for SS1102 and so it cannot be included in the top panel in Fig.~\ref{colors}. The dashed lines are boundaries used by Cieza et al. (2010) to separate the transition discs from the ``full'' discs with no inner holes. The colors for 2M1139 are similar to the sources classified as debris discs in these regions (black points), while 2M1207 and SS1102 lie among the Class II sources. Transition discs with flattened disc structures due to grain growth/dust settling show redder [3.6]-[24]  and [5.8]-[8] colors than the debris discs. While 2M1207 and SS1102 lie to the right of the transition disc boundary, their colors are also similar to discs dominated by grain growth/dust settling. This is consistent with the flat 10 and 20$\mu$m silicate emission features observed for both 2M1207 and SS1102 (Riaz \& Gizis 2008), indicating grain growth to sizes $>$2$\mu$m followed by dust settling to the disc midplane. We can expect weaker excess in the [3.6]-[24] or [5.8]-[8] color for a $\sim$10 Myr old disc compared to the younger $\sim$1 Myr Class II sources in Taurus, most of which are still dominated by sub-micron sized dust grains (Riaz 2009; Sargent et al. 2009). 





There are four widely discussed mechanisms to explain the formation of inner opacity holes in transition discs. The process of giant planet formation would require a large disc mass ($>$ 1 $M_{Jup}$) and therefore can be rejected for 2M1139. Tidal truncation due to a close companion may also not be applicable for 2M1139, as Chauvin et al. (2010) have found no secondaries for this object at projected separations of $\sim$5--240 AU. However, given an inner hole size of $\sim$1 AU for 2M1139, we cannot rule out the possibility of a tight binary surrounded by a circumbinary debris disc. We note that there is no report that 2M1139 is a close double in the limited set of published high resolution spectra. Inner disc evacuation due to significant grain growth and dust settling is expected to exacerbate accretion, resulting in an actively accreting disc with an inner opacity hole (e.g., Cieza et al. 2010). This mechanism would also not be applicable for 2M1139. The transition from a primordial to a debris stage could take place due to the photoevaporation mechanism, which can set in once the accretion rate drops below the photoevaporative wind rate. As soon as photoevaporation sets in, the remaining gas in the disc is removed very rapidly ($<<$ 1 Myr), and what remains is a gas-poor debris disc (e.g., Alexander \& Armitage 2006). For discs that are still in the process of photoevaporation, we can expect weak or negligible accretion but larger fractional disc luminosities than 2M1139, of the order of 10$^{-2}$ (Fig.~\ref{turnoff}). The much lower $L_{D}/L_{*}$ for 2M1139 thus suggests that it has already made a transition to the debris phase. Interestingly, while the 24$\mu$m excess for 2M1139 is larger than that observed for the debris discs TWA 7 and 13 (Fig.~\ref{excess}), it remains undetected at 70$\mu$m and beyond. This is unlike TWA 7 and 13 which show strong excesses at 70$\mu$m (Low et al. 2005). This suggests an absence of cold dust material for 2M1139, and perhaps the outer disc region has been dissipated completely. We cannot, however, conclude that there is no cold dust material surrounding 2M1139, or that this is a case of inner disc clearing due to photoevaporation. There is a limit to the fractional disc luminosity that can be reached with a certain exposure time. We estimate that we can detect an AU Mic like disc ($L_{D}/L_{*}$ $\sim$ 6E-4; Liu et al. 2004) with the $\sim$1 hour total exposure time used for 2M1139. The fractional disc luminosity as estimated from the 2M1139 SED is of the order of $\sim$ 10$^{-5}$, which is well below the detection limit that can be obtained with the requested setup.   


There are results suggesting two distinct populations in the TWA, as evidenced by a bimodal distribution in the rotation periods (Lawson \& Crause 2005) and the presence of warm (T$\ga$ 100 K) dust (Low et al. 2005; Weinberger et al. 2004). Low et al. found negligible amounts of warm dust around 20 out of their 24 TWA targets, while the other four display strong excess emission at 24 $\mu$m (Fig.~\ref{excess}). Lawson \& Crause have found the projected rotational velocities, {\it v} sin {\it i}, for the TWA 1-13 group led by TW Hya to be 3--10 km s$^{-1}$, while the velocities for the stars in the TWA 14-19 group are $\sim$20--50 km s$^{-1}$. These authors have suggested that the former group might be younger than the latter by $\sim$8 Myr (8-10 Myr versus $\sim$17 Myr). 2M1139 has a {\it v} sin {\it i} of 25 km$^{-1}$, which is a factor of $\sim$2 higher than 2M1207 (13 km s$^{-1}$; Scholz et al. 2005). This brown dwarf thus could be slightly older than 2M1207, which could explain its advanced evolutionary stage of being a debris disc, compared to the primordial colors observed for 2M1207 and SS1102. We note that a variety in SED morphologies is observed at any given age (e.g., Hartmann et al. 2005), and the presence of a debris disc does not necessarily imply an older age.




2M1207 is known to have a $\sim$5$M_{Jup}$ secondary that lies at a projected separation of $\sim$55 AU (Chauvin et al.\ 2004). The best model fit for 2M1207 is obtained for an outer disc radius of 50 AU, which suggests tidal truncation of the disc by the massive companion. Such a tidal truncation could be responsible for the sub-mm non-detection of this disc. However, we cannot rule out the non-detection for both 2M1207 and SS1102 in the SPIRE bands due to limitations by the confusion noise. We have discussed in Riaz et al. (2012a) the possibility of the formation of the $\sim$5$M_{Jup}$ secondary in the 2M1207 disc via disc fragmentation. Our argument is that even a very low mass disc could produce a fragment of $\sim$0.035$M_{Jup}$, which can then grow over time to form a 5 $M_{Jup}$ mass object. The main requirement for such a case is for the initial mass of the disc to be higher than its current estimate (at least 10--20 $M_{Jup}$). An upper limit on the disc mass of $\sim$0.1 $M_{Jup}$ as obtained from modeling of 2M1207 SED thus does not rule out disc fragmentation, since the system is relatively old and we have not observed it during its early stages when fragmentation could have occurred ($<$0.1 Myr). This system can rather be considered as a wide very low-mass binary than a planetary system, and some alternative formation mechanisms, such as turbulent fragmentation of a core, have been discussed in Riaz et al. (2012a).

\begin{figure*}
\centering
  \includegraphics[width=12cm]{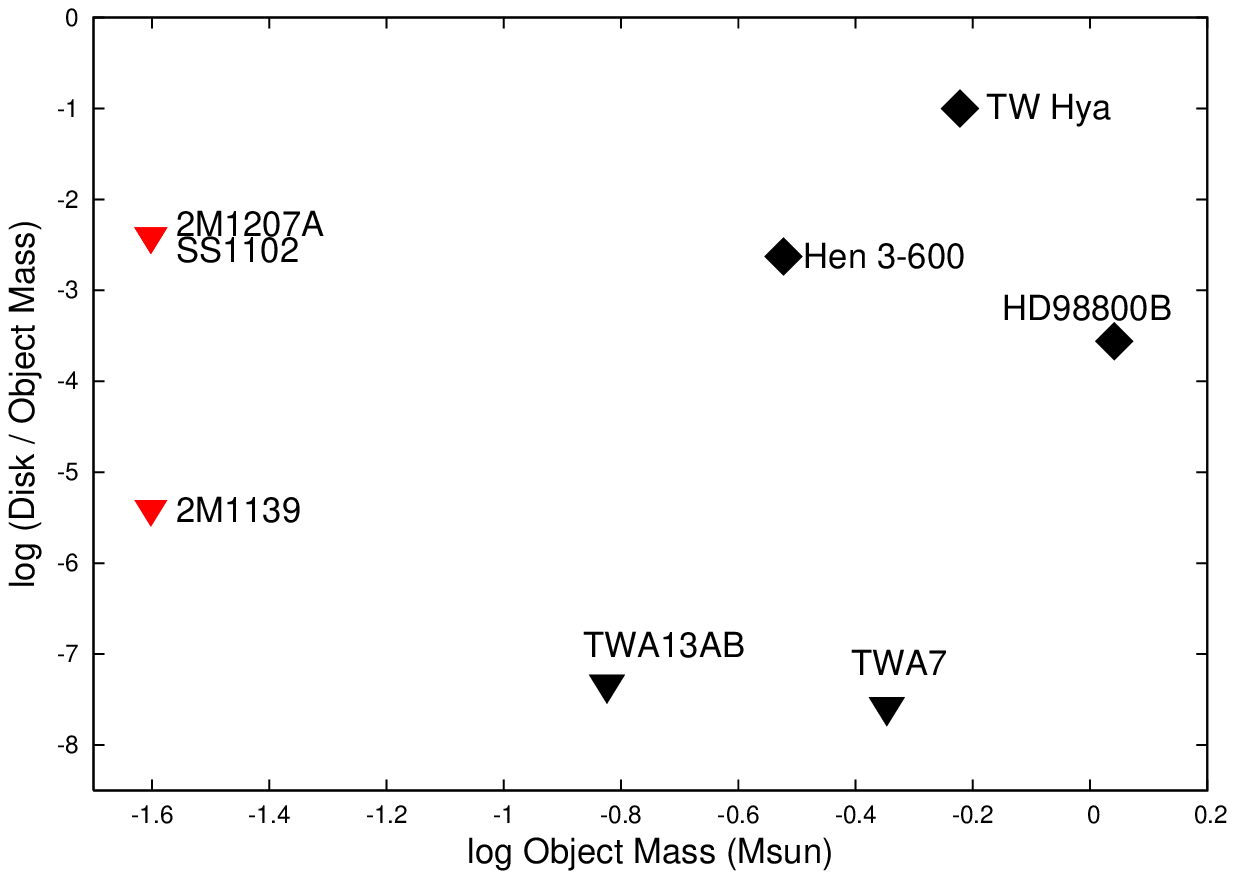} \vspace{1cm}
  \includegraphics[width=12cm]{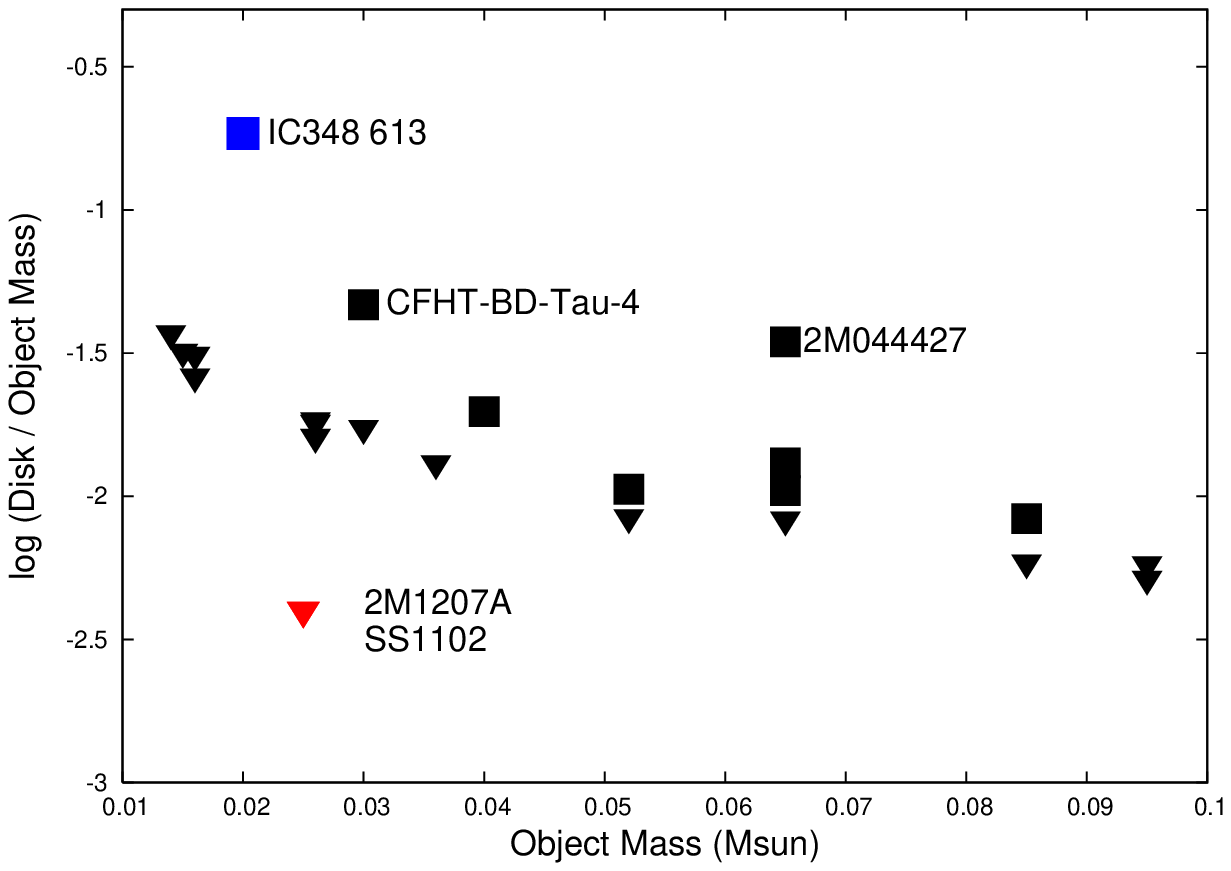}  
 \caption{{\it Top}- (a): Relative disc masses vs. object mass for TWA discs. {\it Bottom}: (b) A comparison of TWA relative disc masses with other known brown dwarf discs in Taurus and IC 348. Upper limits are shown as bold triangles.  }
  \label{discmass}
\end{figure*}

Figure~\ref{discmass}a compares the relative disc mass upper limits for 2M1207, SS1102 and 2M1139 with the other disc sources in the TWA (using data from Calvet et al.\ 2002, Low et al.\ 2005 and Andrews et al.\ 2010). TW Hya is known to be a strong accretor, with $\dot{M}$ of 4 x 10$^{-10} M_{\sun}$ $yr^{-1}$ (Muzerolle et al.\ 2000). Hen 3-600 is a weaker accreting source (0.5x10$^{-10} M_{\sun}$ $yr^{-1}$), while the other disc sources in TWA are non-accretors (e.g., Muzerolle et al.\ 2000). Among the brown dwarfs, 2M1207 shows variable accretion activity of 0.8--1.6 x 10$^{-10} M_{\sun}$ $yr^{-1}$ (Scholz et al.\ 2005; Stelzer et al.\ 2007), while SS1102 and 2M1139 are non-accretors ($<$ 10$^{-11}$ $M_{\sun}$ $yr^{-1}$; Scholz et al.\ 2005). The disc mass limits for 2M1207 and SS1102 are comparable to the weak/non-accretors in the TWA, while 2M1139 is more similar to the other two transition discs TWA 7 and 13. Transition discs are found to have much lower disc masses and accretion rate than non-transition disc sources (e.g., Cieza et al. 2010). Both accreting/non-accreting objects, or sources with or without mid-IR excess are found to have an equal likelihood of a sub-mm detection (e.g., Lee et al. 2011). The lack of detection in the SPIRE bands for 2M1207 and SS1102 could be due to a paucity of millimeter sized dust grains in these discs. Fig.~\ref{discmass}b compares the relative disc masses for the Taurus, IC348 and TWA brown dwarfs, thus covering an age range of $\sim$1-10 Myr. The Taurus and IC 348 disc mass measurements are from Scholz et al.\ (2006) and Klein et al.\ (2003), respectively. The TWA discs have relative masses smaller than nearly all brown dwarf discs in Taurus, thus suggesting that the disc mass may be dependent on the age of the system. For higher mass stars ($\sim$0.1-1 $M_{\sun}$) in the Taurus, Ophiuchus and IC 348 clusters, the disc mass distribution for the relatively older 2-3 Myr clusters is found to be shifted by a factor of $\sim$20 to lower masses, compared to the distribution for the $\sim$1 Myr old clusters (Lee et al. 2011). For the case of the brown dwarfs, most of the data points in Fig.~\ref{discmass}b for object masses $<$ 0.03 $M_{\sun}$ are upper limits. The non-detection of these faint objects in the sub-mm/mm bands could be due to low sensitivity of the observations. To probe any age dependence of the disc mass would require larger samples of brown dwarfs across a wide range in ages.



\section{Summary}

We report a non-detection of SS1102 and 2M1207 in the SPIRE and PACS 160$\mu$m bands, and a non-detection of 2M1139 in the PACS 70 and 160$\mu$m bands. 2M1139 likely possesses a warm debris disc, as indicated by its 24$\mu$m excess emission. We estimate an upper limit to the disc mass for SS1102 and 2M1207 of 0.1 $M_{Jup}$, which places them among the weaker brown dwarf discs in the $\sim$1 Myr old Taurus region, and suggests a decline in the brown dwarf disc mass with the age of the system.  


\onecolumn
\begin{table*}
\centering
\begin{minipage}{10cm}
\caption{Observations}
\begin{tabular}{cccccccccc}
\hline\hline

Band\footnote{Fluxes in [mJy]. References: 1 -- DENIS; 2 -- 2MASS; 3 -- Riaz et al. (2006); 4 -- Riaz \& Gizis (2008); 5 -- Jayawardhana et. al. (2003); 6 -- Sterzik et. al. (2004); 7 -- Harvey et al. (2012); 8 -- This work.} & 2M1207 [Ref] & SS1102\footnote{The fluxes at 5.8, 8 and 24$\mu$m are from the {\it Spitzer}/IRS spectrum.} [Ref]  & 2M1139 [Ref]  \\ \hline

{\it I} & 1.13$\pm$0.1 [1]  & 0.83$\pm$0.1 [1]  & 1.17$\pm$0.1 [1] \\
{\it J} &  10.10$\pm$0.89 [2]  & 9.75$\pm$1.0 [2]  & 13.43$\pm$0.3 [2] \\
{\it H} &  11.35$\pm$1.0 [2]  & 11.7$\pm$1.0 [2]  & 16.3$\pm$0.3 [2] \\
{\it K} & 11.12$\pm$0.98 [2]  & 11.72$\pm$1.0 [2]  & 16.7$\pm$0.3 [2] \\
3.6$\mu$m &  8.49$\pm$0.32 [3]  & -- & 12.08$\pm$0.7 [3] \\
4.5$\mu$m & 7.15$\pm$0.26 [3]  & -- & 8.5$\pm$0.6 [3] \\
5.8 $\mu$m  &  6.36$\pm$0.06 [3]   & 7.6$\pm$0.2 [6]  & 5.98$\pm$0.6 [3] \\
8$\mu$m  &  5.74$\pm$0.21 [3]   & 6.74$\pm$0.2 [6]  & 3.88$\pm$0.5 [3] \\
3.8$\mu$m &  7.00$\pm$0.6 [5]  & -- & 10.23$\pm$1 [5] \\
8.7$\mu$m &  5.60$\pm$1 [6]  & -- & -- \\
10.4$\mu$m &  7.50$\pm$1 [6]   & -- & -- \\ 
24$\mu$m  &  4.32$\pm$0.03 [3]   & 3.96$\pm$0.1 [3]  & 1.10$\pm$0.2 [3] \\
70$\mu$m & 9$\pm$4 [4]  & 9.4$\pm$0.6 [7]  & $<$0.2 [8] \\
160$\mu$m & $<$0.2 [8] & $<$0.2  [8]  & $<$0.075 [8] \\
250$\mu$m & $<$5.2 [8] & $<$1.5 [8] & -- \\
350$\mu$m &$<$5 [8] & $<$1 [8] & -- \\
500$\mu$m & $<$16 [8] & $<$1 [8] & -- \\ 

\hline
\end{tabular}
\label{fluxes}
\end{minipage}
\end{table*}

\begin{table}
\centering
\begin{minipage}{10cm}
\caption{Disk parameters}
\begin{tabular}{cccc}
\hline\hline
Parameter  & 2M1207 & SS1102 & 2M1139 \\ \hline

$\beta$ & 1.2$\pm$0.01 & 1.12$\pm$0.01 & 1.0$\pm$0.01 \\
$h_{0}$ & 0.3$\pm$0.1 & 0.25$\pm$0.1 & 0.1$\pm$0.1 \\
$R_{min}$ & 1 $R_{sub}$ ($\sim$3 $R_{*}$) & 1 $R_{sub}$ & 7 $R_{sub}$ \\
$R_{max}$ & 50 -- 100 AU & 50 -- 100 AU &  100 AU \\
$M_{disc}$ & $<$0.1 $M_{Jup}$ & $<$0.1 $M_{Jup}$ & $<$1E-7 $M_{Jup}$ \\
{\it i} & 57$\degr$ -- 69$\degr$ & 53$\degr$ -- 63$\degr$ & 50$\degr$ \\

\hline
\end{tabular}
\label{disc}
\end{minipage}
\end{table}

\end{document}